\documentclass[letterpaper,journal]{IEEEtran}
\usepackage{amsmath,amsfonts,amssymb}
\usepackage{amsthm}
\usepackage{algorithm}
\usepackage{algpseudocode}
\usepackage{algorithmicx}
\usepackage{array}
\usepackage[caption=false,font=footnotesize,labelfont=rm,textfont=rm]{subfig}
\usepackage{textcomp}
\usepackage{stfloats}
\usepackage{cancel}
\usepackage{url}
\usepackage{multirow}
\usepackage{makecell}
\usepackage{verbatim}
\usepackage{diagbox}
\usepackage{booktabs}
\usepackage{graphicx}
\usepackage{cite}
\usepackage{hyperref}
\hypersetup{hidelinks}
\usepackage{mathrsfs}
\usepackage{bbm}
\usepackage{newtxtext}

\newcommand{\systemname}{\textit{PriorityFEC}}

\providecommand{\safeincludegraphics}[2][]{%
\IfFileExists{#2}{\includegraphics[#1]{#2}}{%
\fbox{\parbox[c][1.22in][c]{0.94\linewidth}{\centering
Figure placeholder: \texttt{\detokenize{#2}}}}}}

\begin{document}

\title{PriorityFEC: Deadline-Aware Streaming FEC for KV-Cache Handoff in Edge Multi-Agent Systems}
\author{Han Shen, Qifu Tyler Sun,~\IEEEmembership{Member,~IEEE,}  Fanyang Meng,~\IEEEmembership{Member,~IEEE}, Ye Wang and Yongsheng Liang
\thanks{\textit{Corresponding authors: Qifu Tyler Sun, Fanyang Meng}.}}

\maketitle

\begin{abstract}
Collaborative edge large language model (LLM) systems often require repeated prefill computation when task contexts are transferred between agents. Reusing previously computed key--value (KV) caches can eliminate this overhead, but requires KV-cache delivery before the receiving agent begins decoding. Under strict deadlines and unreliable edge links, complete KV-cache transfer may be infeasible, making the timely delivery of useful KV information critical. This paper proposes PriorityFEC, a utility-aware KV-cache handoff scheme that combines prioritized source transmission with adaptive loss recovery. PriorityFEC first organizes KV blocks into a utility-aware source queue to prioritize important information. It then employs ACK-sliding Streaming FEC and a lightweight Double Deep Q-Network (Double-DQN) scheduler to dynamically balance source delivery and repair transmission under deadline constraints. We implement PriorityFEC in a distributed edge LLM-serving prototype and evaluate it using Qwen2.5-7B-Instruct and Llama-3.1-8B-Instruct across multiple workloads. Experimental results demonstrate that PriorityFEC preserves more useful KV information and achieves substantially higher task accuracy than competing transport schemes, particularly when the deadline is insufficient to deliver the complete KV cache.
\end{abstract}



\section{Introduction}
\label{sec:introduction}

Large language models (LLMs) are increasingly used in collaborative edge applications, where multiple agents work together to complete long-running tasks \cite{edgeshard,feng20262}. For example, in collaborative robotics, a robot may continuously interact with its environment, observe events, and receive task instructions. When another robot joins or takes over the task, it needs information from these previous interactions to make appropriate decisions. However, past events cannot always be observed again, and their complete records may not be available to the new robot. Without such information, the new robot may have difficulty understanding the current task. Therefore, efficiently sharing previously acquired information is important for collaboration among LLM agents.

KV-cache reuse provides a promising solution to this problem. During the prefill stage, an LLM processes the input context and generates a key--value (KV) cache for subsequent token generation. When a task is handed off, the upstream agent can transfer its KV cache to a receiving agent running a compatible model. The receiving agent can then reuse the previously processed context without repeating the entire prefill computation. 
This reduces redundant computation and shortens
the time-to-first-token (TTFT), enabling the
receiving agent to resume decoding on time to
meet the latency requirements of the task.

\begin{figure}[t]
    \centering
    \includegraphics[width=0.485\textwidth]
    {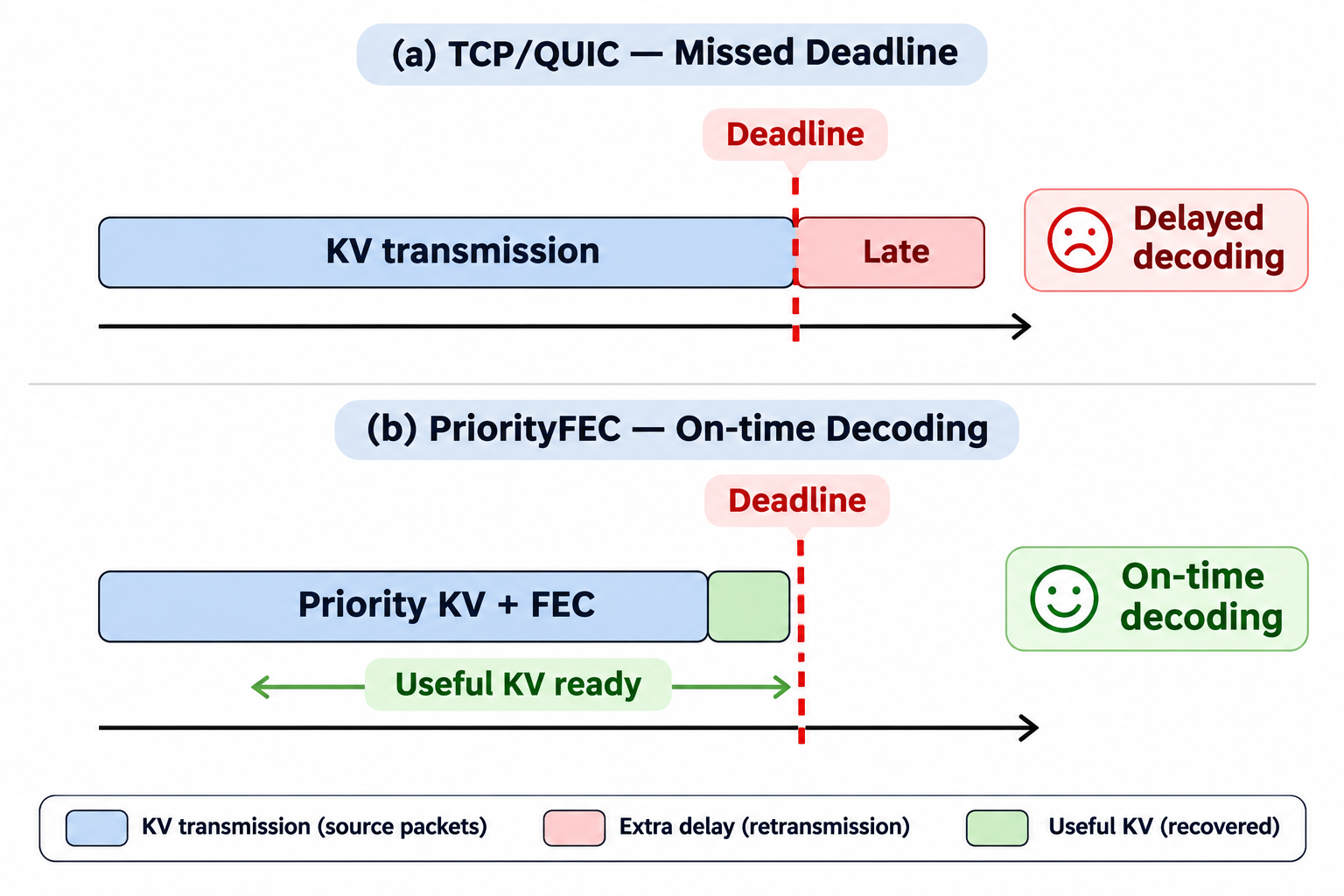}
    \caption{Motivation for deadline-constrained KV-cache handoff.
    (a) Conventional reliable transport relies on retransmissions,
    which may delay KV-cache recovery beyond the decode-start deadline.
    (b) \systemname{} prioritizes high-utility KV blocks and employs
    FEC-based loss recovery, enabling the receiving agent to start
    decoding on time with useful KV information.}
    \label{fig:transport-motivation}
\end{figure}

However, transferring KV caches between edge agents introduces significant communication challenges. In collaborative edge systems, agents may communicate over bandwidth-constrained and unreliable wireless links. Long interaction histories can generate large KV caches, resulting in considerable transmission delays. Moreover, wireless interference, mobility, and handovers may cause bursty packet losses, further delaying KV-cache delivery. Consequently, the complete KV cache may not arrive before the receiving agent must resume inference, making timely KV-cache handoff particularly challenging~\cite{chen2025kvdirect}.

Existing KV-cache systems primarily focus on cache
storage, reuse, compression, and efficient data
movement in data centers and GPU clusters
\cite{liu2025lmcache,liu2024cachegen}.
However, transferring KV caches across
resource-constrained and unreliable edge links
introduces additional challenges, particularly
when the receiving agent must resume decoding
within a strict deadline. In such scenarios,
limited bandwidth and packet losses may prevent
the complete KV cache from being delivered in
time. Waiting for all missing data to arrive
would delay task execution, whereas starting
decoding with incomplete KV information may
degrade inference quality. This creates a
fundamental tradeoff between timely task
resumption and the completeness of the
KV cache. Rather than requiring
the eventual delivery of the entire KV cache,
our objective is therefore to maximize the
utility of the KV information available when
the receiving agent must begin decoding. 

Achieving this objective requires a transport
mechanism that can recover packet losses without
consuming excessive transmission time, while
allowing useful KV information to arrive before
the deadline. However, conventional transport
mechanisms are not specifically designed for
this requirement. As illustrated in
Fig.~\ref{fig:transport-motivation}, reliable
protocols such as TCP~\cite{hunt2002tcp} and
QUIC~\cite{langley2017quic} recover lost data
through retransmissions, which may delay the
availability of useful KV information.
Conversely, unreliable transport such as UDP
avoids retransmission delays but provides no
loss recovery, potentially leaving important
KV information unavailable at decoding time.
Packet-level forward erasure correction (FEC)
offers an alternative by introducing repair
packets that enable the receiver to recover
lost source packets without waiting for
retransmissions~\cite{michel2019quicfec}.

Two common FEC approaches are block coding and
streaming coding. Block codes, such as
Reed--Solomon~\cite{wicker1999reed} and
Raptor codes~\cite{shokrollahi2006raptor},
generate redundancy from predefined groups of
source packets. Although effective for erasure
recovery, block-based operation may introduce
additional recovery delays when coding blocks
remain incomplete. In contrast, Streaming FEC
\cite{li2023low} generates repair packets over
an evolving coding window, allowing redundancy
to protect source packets continuously as
transmission progresses. This flexibility is
particularly attractive for deadline-constrained
KV-cache delivery, where the sender must balance
the transmission of new KV information against
the recovery of previously lost packets.
Therefore, we adopt Streaming FEC as the basis
for the proposed deadline-aware KV-cache
transport mechanism.

However, the flexibility of Streaming FEC does not eliminate the fundamental tradeoff between source transmission and loss recovery under strict deadlines. Sending more repair packets improves the recoverability of transmitted KV cache but reduces the opportunities to deliver additional useful KV information. While fixed-rate FEC cannot adapt redundancy to changing network conditions, existing feedback-adaptive schemes primarily rely on packet losses and decoding progress without explicitly accounting for KV utility. Consequently, effective repair scheduling must consider not only the recovery needs of transmitted KV blocks but also the utility of those awaiting transmission.


The above observations expose two coupled challenges for
deadline-constrained KV-cache handoff.
\begin{itemize}
    \item \textbf{\emph{First, which KV blocks should receive the earliest transmission
    opportunities?}}
    When the complete KV cache cannot be delivered before decode startup,
    different source positions have different opportunities to become
    available by the deadline. Treating all KV blocks uniformly can
    therefore waste favorable source positions on modeling state with limited
    downstream utility.

\item \textbf{\emph{Second, how should transmission opportunities be
allocated between new KV packets and FEC repair packets
under a strict deadline?}}
Sending additional source packets increases the amount
of potentially useful KV cache delivered to the receiver,
whereas sending repair packets improves the recoverability
of previously transmitted KV information.
However, both consume the same limited transmission
opportunities. The optimal tradeoff therefore depends
not only on network losses and decoding progress,
but also on the utility of KV blocks at risk and
those awaiting transmission.

\end{itemize}

Motivated by these challenges, we develop
\systemname{} around two complementary mechanisms. First, a
\emph{utility-aware source ordering} maps model-derived KV utility to
utility blocks and places high-utility blocks earlier in
the transmission stream. Second, a \emph{feedback-adaptive Streaming FEC}
mechanism protects the stream by adjusting repair
transmissions according to the TTFT deadline.

These two mechanisms address different aspects of deadline-constrained
KV handoff: utility-aware ordering determines \emph{which} KV caches
receive earlier delivery opportunities, whereas adaptive Streaming FEC
determines \emph{when} recovery opportunities should be introduced.
Together, they improve the utility of the reconstructed KV cache before
the TTFT deadline without requiring packet-level retransmissions.

The main contributions of this work are summarized as follows.

\begin{itemize}

    \item \textbf{Deadline-constrained KV-cache handoff.}
    We investigate KV-cache handoff in collaborative edge
    LLM systems under strict decoding deadlines, where
    the available transmission time may be insufficient
    to deliver the complete KV cache even without packet
    losses. Instead of requiring complete data delivery,
    we formulate KV-cache handoff as a deadline-constrained
    optimization problem that maximizes the utility of
    KV cache recovered before decoding begins.
    This formulation provides a basis for balancing
    the transmission of new KV packets against
    the recovery of lost packets.

    \item \textbf{Utility-Aware Source Ordering and Queue Design.}
    We propose PriorityFEC, which maps offline model-derived KV importance scores to utility blocks and organizes them into a utility-aware transmission queue, placing higher-utility blocks at earlier and more favorable source positions for stronger pre-deadline protection.

\item \textbf{Deadline-aware source--repair optimization.}
We formulate the allocation of transmission opportunities
between new KV packets and FEC repair packets as a
finite-horizon stochastic scheduling problem under strict
deadline and repair-budget constraints.
Unlike conventional FEC scheduling that primarily responds
to network losses and decoding deficits, our formulation
explicitly accounts for the utility of KV blocks at risk
and those awaiting transmission.
We develop a lightweight Double Deep Q-Network (Double-DQN)-based scheduler
that jointly considers KV utility, decoding progress,
network feedback, and the remaining transmission budget
to dynamically balance source delivery and loss recovery.
This enables utility-aware repair decisions that aim
to maximize the KV cache available at decode startup.

\end{itemize}

\section{Related Work}
\label{sec:related_work}

\subsection{KV-Cache Reuse and Movement in Distributed LLM Serving}
\label{subsec:related_kv_movement}

Recent studies have explored KV-cache reuse and transfer
to improve the efficiency of distributed LLM serving.
DistServe~\cite{zhong2024distserve} and
Splitwise~\cite{patel2024splitwise} separate the prefill
and decoding stages across different GPU workers,
allowing previously computed KV caches to be reused.
Mooncake~\cite{qin2024mooncake} further introduces
a KV-cache-centric serving architecture to support
efficient KV sharing among workers.

Other studies focus on KV-cache management and
communication efficiency.
PagedAttention~\cite{kwon2023vllm} improves GPU memory
utilization through efficient KV-cache management, while
LMCache~\cite{liu2025lmcache} supports KV-cache reuse,
offloading, and transfer across serving instances.
KVDirect~\cite{chen2025kvdirect} provides tensor-aware
communication for efficient KV transfer.
CacheGen~\cite{liu2024cachegen} reduces transmission
overhead through KV-cache compression, while
KVServe~\cite{liu2026kvserve} adapts compression
to workload and resource requirements.

These studies mainly focus on efficient KV-cache
management, compression, and transfer.
In contrast, our work considers KV-cache handoff
between edge agents over unreliable networks.
When the complete KV cache cannot be delivered
within a strict deadline, our objective is to
prioritize the KV information that is most useful
for subsequent inference.

\subsection{FEC for Deadline-Constrained and Lossy Transport}

Previous studies have investigated reliable inference
over lossy wireless networks, showing that partial
information recovery can support inference under
strict latency constraints
\cite{itahara2021packetloss,itahara2022comtune,hou2025lossadapter}.
FEC provides an effective
means of recovering packet losses without waiting
for retransmissions.
Conventional block FEC divides source packets into
separate coding blocks and generates repair packets
for each block~\cite{ho2006rlnc,wicker1999reed}.
However, repair packets are generally restricted
to their corresponding blocks, limiting recovery
when burst losses exceed the allocated redundancy.
Streaming FEC addresses this limitation by
protecting source packets through overlapping
coding windows
\cite{tournoux2011tetrys,badr2013streaming,li2023low}.
Unrecovered packets can remain in the coding
window and benefit from subsequent repair packets,
making Streaming FEC suitable for bursty-loss
transmission.

Several studies further improve FEC performance
through adaptive redundancy control.
DeepRS~\cite{cheng2020deeprs} predicts packet losses
using an LSTM and adjusts the FEC redundancy ratio,
while RL-AFEC~\cite{chen2022rlafec} employs
reinforcement learning to select Reed--Solomon
redundancy for real-time video transmission.
QUIC-FEC~\cite{michel2019quicfec} integrates erasure
coding into QUIC, while AC-RLNC~\cite{co2021acrlnc}
adapts coding decisions based on delayed feedback.
FlEC~\cite{michel2023flec} further supports
application-aware reliability control and
repair scheduling in QUIC.
These methods improve loss recovery by adapting
redundancy to network conditions. In our work,
the scheduler must additionally consider the
utility of KV packets and the remaining deadline.
Sending repair packets improves recovery but
also reduces the time available for transmitting
new KV cache.

Unequal error protection (UEP) provides another
approach to improving the delivery of important
information. Priority Encoding Transmission
\cite{albanese1996priority} provides different
recovery guarantees for source components, while
expanding-window fountain codes
\cite{sejdinovic2009expanding} offer stronger
protection to high-priority data.
UEP random linear coding
\cite{vukobratovic2012unequal} uses different
coding windows to protect packets according
to their importance.
Prioritized RLNC~\cite{thomos2015adaptive}
further combines coding and scheduling decisions
for layered data delivery.
Unlike these approaches, which primarily
incorporate data importance into coding windows
or protection policies, \systemname{} uses
model-derived KV utility to determine the
source transmission order and guide adaptive
repair scheduling under a strict deadline.

\section{Utility-Aware Source Ordering for Deadline-Constrained KV Handoff}
\label{sec:priority_queue}
\subsection{Transport-Oriented KV-Utility Validation}
\label{subsec:utility_validation}

\begin{figure}[t]
    \centering
    \includegraphics[
        width=\linewidth,
        trim=5bp 24bp 2bp 9bp,
        clip
    ]{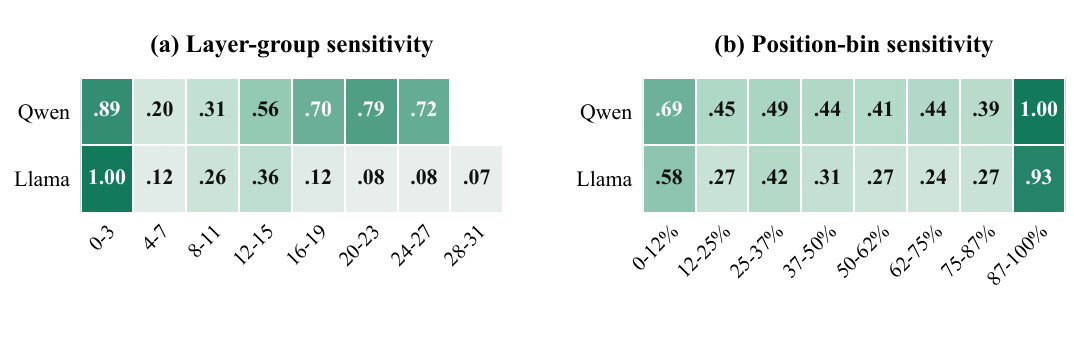}

    \par\vspace{0.25em}

    \includegraphics[
        width=0.72\linewidth,
        trim=1bp 9bp 2bp 7bp,
        clip
    ]{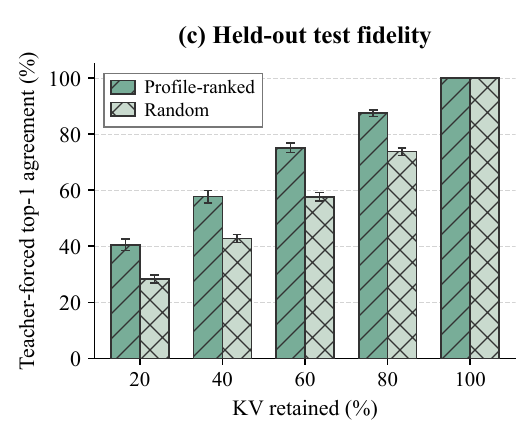}

    \setlength{\abovecaptionskip}{4pt}
    \caption{Non-uniform KV-cache utility.
    (a) Normalized layer-group sensitivity.
    (b) Normalized token-position sensitivity.
    (c) Held-out teacher-forced top-1 agreement under partial
    KV retention. Profile-ranked retention outperforms random
    retention; error bars denote 95\% confidence intervals.}
    \label{fig:kv_utility_characterization}
\end{figure}

Prior studies on KV-cache compression have shown
that different KV regions contribute unequally to model
inference~\cite{h2o,snapkv,pyramidkv}. Our objective is therefore not to introduce a new universal
KV-importance estimator. Instead, we ask a transport-oriented question:
When only part of the KV cache can be made available before decode
startup, does the identity of the recovered KV cache materially affect
model fidelity?

We evaluate Qwen2.5-7B-Instruct~\cite{qwen25} and Llama-3.1-8B-Instruct~\cite{llama3} through
controlled \textit{KV-zeroing} experiments to simulate the packet loss during the KV cache transmission. For each profiling prompt, we first
record the next-token distribution obtained using the full KV cache.
We then zero both key and value tensors associated with one layer group
or one relative-position region and measure the resulting
distributional change. The perturbation scores are averaged over the
profiling set and normalized within each model.

Fig.~\ref{fig:kv_utility_characterization}(a) shows that
KV sensitivity varies considerably across layers.
Both models exhibit high sensitivity in the early layers,
but their sensitivity patterns differ in the remaining
layers. This suggests that a fixed layer-based transmission
order may not be suitable for different LLMs.
Fig.~\ref{fig:kv_utility_characterization}(b) further
shows that KV sensitivity varies across token positions.
In particular, the beginning and end of the input
context are generally more sensitive than the
intermediate positions.

To evaluate whether these sensitivity profiles generalize
to unseen inputs, we conduct experiments on held-out
prompts. For each KV retention ratio, we preserve the
highest-ranked layer--position regions and compare them
with randomly selected regions.
As shown in Fig.~\ref{fig:kv_utility_characterization}(c),
profile-based retention consistently achieves higher
teacher-forced top-1 agreement with full-KV inference.
At a retention ratio of 40\%, the agreement increases
from 44.0\% with random retention to 55.2\% with
profile-based retention.

These results demonstrate that the quality of recovered
KV cache depends not only on the amount of data
available but also on which KV regions are preserved.
This motivates utility-aware KV-cache transmission,
which prioritizes important KV regions to improve
inference quality under strict transmission deadlines.


\subsection{Mapping KV Utility to Transport Blocks}
\label{subsec:utility_mapping}

To incorporate KV importance into packet transmission,
we divide the serialized KV cache into $B$ transport
blocks, indexed by
\begin{equation}
    \mathcal{B} = \{1,\ldots,B\}.
    \label{eq:block_set}
\end{equation}
Each block contains $m$ fixed-size source packets,
\begin{equation}
    \mathcal{P}_b
    =
    \{\mathbf{s}_{b,1},\ldots,\mathbf{s}_{b,m}\},
    \qquad b\in\mathcal{B}.
    \label{eq:block_packet_set}
\end{equation}
The final block is padded if necessary to ensure
equal block sizes, with zero utility assigned to
padding data.

Let $\mathcal{C}$ denote the set of profiled
layer--position cells, and let $\widehat{u}_c\geq 0$
represent the measured KV sensitivity of cell
$c\in\mathcal{C}$.
Since a profiled cell may span multiple packets,
we define $\alpha_{i,c}\in[0,1]$ as the fraction
of cell $c$ contained in source packet $i$, where
\begin{equation}
    \sum_i \alpha_{i,c}=1,
    \qquad \forall c\in\mathcal{C}.
    \label{eq:cell_fraction_conservation}
\end{equation}
The utility of source packet $i$ is then calculated as
\begin{equation}
    v_i
    =
    \sum_{c\in\mathcal{C}}
    \alpha_{i,c}\widehat{u}_c.
    \label{eq:packet_utility}
\end{equation}
Accordingly, the utility of transport block $b$
is obtained by summing the utilities of its
source packets:
\begin{equation}
    U_b
    =
    \sum_{i\in\mathcal{I}_b}v_i,
    \label{eq:block_utility}
\end{equation}
where $\mathcal{I}_b$ denotes the set of source
packet indices belonging to block $b$.

Note that $U_b$ serves as a measure of KV importance
for transmission scheduling rather than an exact
estimate of downstream inference quality.
This is because missing KV regions may jointly
affect the attention computation.
Therefore, our analysis focuses on the utility
of fully recovered transport blocks, while
Sec.~\ref{sec:evaluation} directly evaluates
the impact of partial KV recovery on
downstream task performance.

\subsection{Utility-Aware Source Queue Construction}
\label{subsec:priority_queue}

As described in Sec.~\ref{subsec:utility_mapping},
we divide the KV cache into transport blocks with
different utility values. In collaborative edge LLM
serving, the receiving agent may need to start
decoding under a strict deadline. Due to limited
bandwidth and packet losses, the complete KV cache
may not be available before this deadline.
Under such conditions, the transmission order
becomes particularly important. By prioritizing
high-utility KV blocks, the sender can deliver
the most valuable KV information earlier,
allowing the receiving agent to maintain better
inference quality even when only part of the
KV cache is recovered.

Based on this observation, we construct a
utility-aware source queue by sorting KV blocks
in nonincreasing order of their utilities:
\begin{equation}
U_{[1]}\geq U_{[2]}\geq\cdots\geq U_{[B]},
\label{eq:utility_order}
\end{equation}
where $[j]$ denotes the index of the block
assigned to the $j$-th transmission position.
The source packets within each block retain
their original order and are transmitted
systematically. Since all blocks contain
the same number of equal-sized packets,
the sorting requires no additional
normalization by block size.

Although utility-aware ordering prioritizes
the transmission of important KV blocks,
it does not guarantee their successful recovery
over unreliable links. Packet losses may still
prevent high-utility blocks from being completely
reconstructed before the deadline. Therefore,
we further introduce ACK-sliding Streaming FEC
and adaptive repair scheduling in the following
section to protect transmitted KV information
and improve its timely recovery.

\section{Deadline-Aware Streaming FEC and Repair Scheduling}
\label{sec:adaptive_streaming_fec}

The utility-aware source queue determines which KV blocks should be transmitted first, but cannot by itself recover packet losses over unreliable edge links. 
To address this, we combine the utility-ordered source queue with ACK-sliding Streaming FEC and an adaptive repair scheduler. The queue determines \emph{what} to transmit first, Streaming FEC provides loss recovery, and the scheduler determines \emph{when} repair packets should be transmitted.

\subsection{ACK-Sliding Streaming FEC}
\label{subsec:ack_sliding_streaming}
\begin{figure}[t]
    \centering
    \safeincludegraphics[width=\linewidth]
    {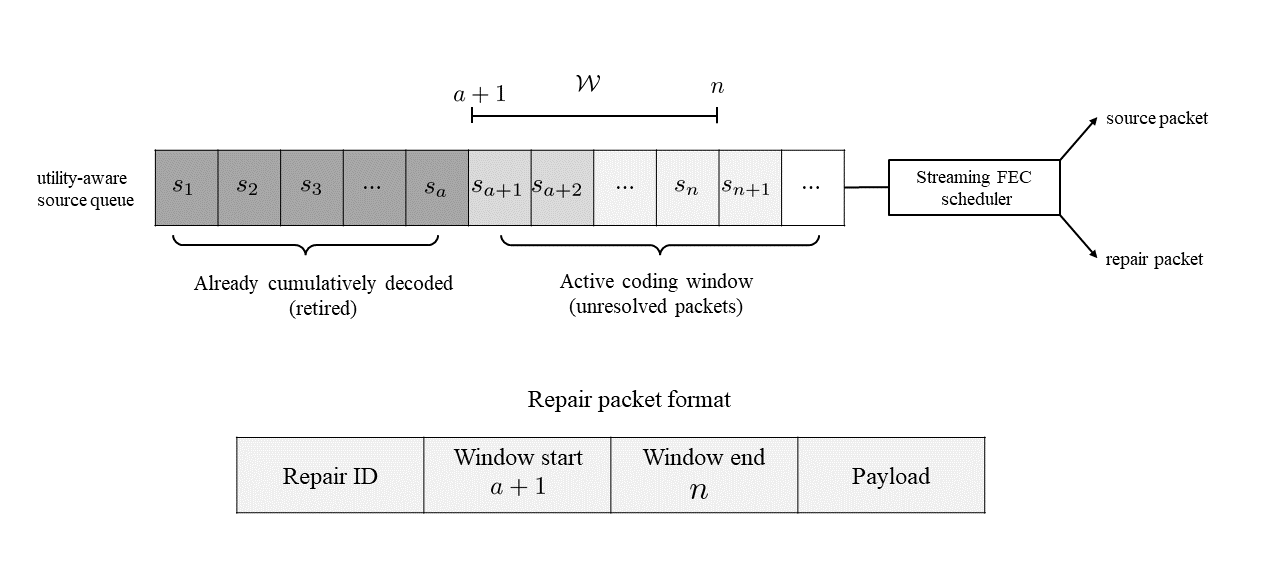}
    \caption{ACK-sliding Streaming FEC. Source packets are transmitted
    systematically, while each repair packet is generated over the
    current unacknowledged source window. Once cumulative decoding
    progresses, the recovered prefix is removed from subsequent coding
    windows; unresolved packets remain protected by future repairs.}
    \label{fig:streaming_fec}
\end{figure}

Let
$\mathbf{s}_1,\mathbf{s}_2,\ldots,\mathbf{s}_N$
denote the equal-size source packets after utility-aware ordering.
Source packets are transmitted sequentially according to this fixed order.

Let $n$ denote the latest transmitted source
index and let $a$ denote the largest source index known by the sender
to have been cumulatively decoded. As shown in Fig.~\ref{fig:streaming_fec}, the active coding window is
\begin{equation}
\mathcal{W}
=
\left\{
\mathbf{s}_{a+1},
\ldots,
\mathbf{s}_{n}
\right\}.
\label{eq:adaptive_active_window}
\end{equation}
Thus, $\mathcal{W}$ contains all transmitted source packets that have
not yet been removed by cumulative ACK progress.

To reduce encoding and decoding complexity, we employ sparse random linear
coding over the active window. Specifically, each repair packet combines
only a subset of the packets in $\mathcal{W}$. Let $K$ denote the maximum
coding degree and define
\begin{equation}
K
=
\min
\left\{
K,
|\mathcal{W}|
\right\}.
\label{eq:sparse_coding_degree}
\end{equation}
For the $\ell$th repair packet generated at decision epoch $t$, a sparse
coding support
$\mathcal{C}_{\ell}\subseteq\mathcal{W}$
is selected such that
\begin{equation}
|\mathcal{C}_{\ell}|
=
K.
\label{eq:sparse_coding_support}
\end{equation}
The corresponding repair packet is
\begin{equation}
\mathbf{r}_{\ell}
=
\sum_{\mathbf{s}_i\in\mathcal{C}_{\ell}}
\alpha_{\ell,i}\mathbf{s}_i,
\qquad
\alpha_{\ell,i}
\in
\mathrm{GF}(2^8)\setminus\{0\},
\label{eq:adaptive_repair_packet}
\end{equation}
where the coding coefficients are independently generated for different
repair packets. In our implementation, the maximum sparse coding degree is
set to $K=48$.

When a source packet is erased, it remains in the active window and can
therefore participate in subsequent sparse repair combinations. Once the
receiver accumulates sufficient independent degrees of freedom (DoFs) to
advance cumulative decoding, the corresponding decoded prefix is reported
through ACK feedback and removed from $\mathcal{W}_t$. Consequently, the
coding window continuously tracks the unresolved portion of the transmitted
source stream.


\textit{Unequal protection property:}
The ACK-sliding mechanism provides an inherent
temporal protection advantage to earlier source
packets, resembling the unequal error protection
property of expanding-window codes.
To illustrate this property, consider two source
packets $\mathbf{s}_i$ and $\mathbf{s}_j$ with
$i<j$.

For the $\ell$th repair packet, let
$a_\ell$ and $n_\ell$ denote the cumulative
ACK boundary and the latest transmitted source
index, respectively. The coding window has size
$W_\ell=n_\ell-a_\ell$, and the coding degree is
$K_\ell=\min\{K_{\max},W_\ell\}$.
Since the coding support is uniformly sampled
without replacement, the probability that
$\mathbf{s}_i$ is included in this repair is
\begin{equation}
p_{i,\ell}
=
\mathbf{1}\{a_\ell<i\leq n_\ell\}
\frac{K_\ell}{W_\ell}.
\label{eq:repair_inclusion_probability}
\end{equation}

As long as the earlier packet $\mathbf{s}_i$
remains unacknowledged, i.e., $a_\ell<i<j$,
the difference in their inclusion probabilities
satisfies
\begin{equation}
\begin{aligned}
p_{i,\ell}-p_{j,\ell}
&=
\frac{K_\ell}{W_\ell}
\mathbf{1}\{i\leq n_\ell<j\}\\
&\geq 0.
\end{aligned}
\label{eq:unequal_repair_protection}
\end{equation}

Therefore, before $s_i$ is cumulatively acknowledged,
its probability of being included in a given repair
packet is no smaller than that of a later source
packet $s_j$, where $i<j$. 

Since high-utility KV blocks are transmitted
first, this property provides them with
additional opportunities for repair under a
limited deadline. Unlike conventional
expanding-window coding, ACK-sliding Streaming
FEC removes the acknowledged prefix to avoid
unnecessary protection of recovered data.

\subsection{Adaptive Repair-Scheduling Problem}
\label{subsec:repair_problem}
Due to limited bandwidth and a strict decode-start
deadline, only a finite number of packets can be
transmitted, and the available time may be insufficient
to deliver even the complete KV cache.
Consequently, each transmission opportunity becomes
particularly valuable. Transmitting additional source
packets allows the receiver to obtain new KV information,
whereas transmitting repair packets helps recover
previously lost high-utility KV packets.
Since both compete for the same limited transmission
opportunities, a key question arises: \textit{how should
the sender balance new source delivery and loss
recovery to maximize the KV utility available
at the decode-start deadline?}

Let $T^{\max}$ denote the maximum number of packet
transmissions allowed before the decode-start deadline,
and let $B^{\mathrm{rep}}$ denote the maximum number of
repair packets that can be transmitted. 

The sender selects
one of the following actions:
\begin{equation}
\mathcal{A}
=
\{a^{\mathrm{s}}\}
\cup
\{a^{\mathrm{r}}(k):k\in\mathcal{K}\},
\label{eq:repair_action_set}
\end{equation}
where $a^{\mathrm{s}}$ denotes that sender transmits $a^{\mathrm{s}}$ source KV packets,
and $a^{\mathrm{r}}(k)$ transmits $k$ repair packets,
with $\mathcal{K}$ denoting the set of available repair
burst sizes. \textbf{At each decision epoch $t$, the scheduler selects either source transmission or a repair burst of a specified size. Once the selected batch of packets has been transmitted, the scheduler proceeds to the next decision epoch $t+1$.}

The objective is to maximize the utility of KV blocks
successfully recovered at the receiver before decoding
begins. The normalized recovered KV utility is defined as
\begin{equation}
J_t^{\mathrm{KV}}
=
\frac{
\sum_{b\in\mathcal{B}} U_b Z_b(t)
}{
\sum_{b\in\mathcal{B}} U_b
}.
\label{eq:normalized_completed_utility}
\end{equation}
where $\mathcal{B}$ denote the set of KV blocks,
$U_b$ the utility of block $b$, and
$Z_b(t)\in\{0,1\}$ indicate whether block $b$ has
been completely recovered by decision epoch $t$.

Accordingly, we formulate the repair-scheduling
problem as

\begin{subequations}
\label{eq:repair_problem}
\begin{align}
\max_{\pi}\quad
& \mathbb{E}_{\pi}
\left[J_{\mathrm{end}}^{\mathrm{KV}}\right],
\\
\mathrm{s.t.}\quad
& u_t\in\mathcal{A}_t,
\\
& N^{\mathrm{rep}}\leq B^{\mathrm{rep}},
\\
& N^{\mathrm{src}}+N^{\mathrm{rep}}\leq T^{\max},
\end{align}
\end{subequations}
where $\pi$ denotes the scheduling policy that
selects an action $u_t\in\mathcal{A}_t$ based on
the current state $\mathbf{x}_t$.
$N^{\mathrm{src}}$ and $N^{\mathrm{rep}}$ denote
the total numbers of source and repair packets
transmitted before the deadline, respectively.

\subsection{Markov Decision Process Formulation}
\label{subsec:repair_mdp}

The source--repair tradeoff depends on the remaining
transmission budget, decoding progress, and KV-block
utility. We therefore formulate adaptive repair
scheduling as a finite-horizon Markov decision process
(MDP) $\langle\mathcal{S},\mathcal{A},P,\mathcal{R}\rangle$,
where $\mathcal{S}$ denotes the state space,
$\mathcal{A}$ is the action space defined in
\eqref{eq:repair_action_set}, $P$ describes the
state transitions, and $\mathcal{R}$ is the reward
function. 

\subsubsection{State Representation}

We construct a compact state representation based
on transmission progress, network conditions,
decoding status, and KV-block utility. Specifically,
the scheduler uses a 15-dimensional state vector
\begin{equation}
\begin{aligned}
\mathbf{x}_t =
[
&
b_t,\,
\tau_t,\,
p_t^{\mathrm{src}},\,
s_t^{\mathrm{ddl}},\,
\hat{\epsilon}_t,\,
\beta_t,\,
\delta_t,\,
\chi_t,
\\
&
w_t,\,
w_t^{\mathrm{fb}},\,
h_t^{\mathrm{ack}},\,
g_t,\,
\bar U_t,\,
\bar U_t^{+},\,
p_t^{\mathrm{eff}}
].
\end{aligned}
\label{eq:rl_state}
\end{equation}
The features are summarized in
Table~\ref{tab:rl_state}.

\begin{table}[t]
\centering
\caption{State Features for Adaptive Repair Scheduling}
\label{tab:rl_state}
\footnotesize
\renewcommand{\arraystretch}{1.1}
\begin{tabular}{@{}p{0.25\columnwidth}p{0.69\columnwidth}@{}}
\hline
\textbf{Feature} & \textbf{Description} \\
\hline

\multicolumn{2}{@{}l}{\textbf{Transmission resources}} \\
$b_t$ & Remaining repair budget. \\
$\tau_t$ & Remaining time before the deadline. \\
$p_t^{\mathrm{src}}$ & Source-transmission pressure. \\
$s_t^{\mathrm{ddl}}$ & Deadline-related source-delivery pressure. \\

\multicolumn{2}{@{}l}{\textbf{Channel conditions}} \\
$\hat{\epsilon}_t$ & Estimated packet-loss rate. \\
$\beta_t$ & Estimated burst-loss tendency. \\

\multicolumn{2}{@{}l}{\textbf{Decoding and feedback status}} \\
$\delta_t$ & Normalized DoF deficit. \\
$\chi_t$ & Severity of unresolved decoding deficit. \\
$w_t$ & Unacknowledged coding-window size. \\
$w_t^{\mathrm{fb}}$ & Feedback-based window information. \\
$h_t^{\mathrm{ack}}$ & Lack of recent cumulative ACK progress. \\

\multicolumn{2}{@{}l}{\textbf{KV-block information}} \\
$g_t$ & Current KV-block transmission progress. \\
$\bar U_t$ & Utility of the current KV block. \\
$\bar U_t^{+}$ & Utility of upcoming KV blocks. \\

\multicolumn{2}{@{}l}{\textbf{Repair effectiveness}} \\
$p_t^{\mathrm{eff}}$ & Estimated repair effectiveness. \\
\hline
\end{tabular}
\end{table}

A key feature is the degree-of-freedom (DoF)
deficit, which measures the amount of additional
independent information required for decoding.
Let $W_t$ denote the current unacknowledged
coding-window size and $D_t$ the number of
independent DoFs available for this window.
The deficit is defined as
\begin{equation}
d_t=[W_t-D_t]^+,
\label{eq:dof_deficit}
\end{equation}
where $[x]^+=\max(x,0)$.
A larger deficit indicates a greater need for
repair information, and its normalized value
is represented by $\delta_t$.

\textbf{Since receiver-side decoding information is
available only through delayed feedback, the
scheduler estimates the decoding-related features
using transmission history and the latest ACKs.}


\subsubsection{Action and Feasibility}

At each decision epoch, the scheduler determines
whether to transmit new source packets or generate
repair packets for loss recovery. The source action
$a^{\mathrm{s}}$ follows the utility-based transmission
order. In contrast,
a repair action $a^{\mathrm{r}}(k)$ generates a burst
of $k$ repair packets for the current unresolved
coding window.

Since both actions consume transmission opportunities,
repair transmission may reduce the time available
for delivering new KV packets. The scheduler must
therefore balance the benefit of recovering missing
packets against the opportunity cost of delaying
source delivery.

To ensure that scheduling decisions satisfy the
transmission constraints, we employ a state-dependent
action mask. Specifically, a repair action
$a^{\mathrm{r}}(k)$ is feasible only when
\begin{equation}
\begin{aligned}
d_t &> 0,\\
B_t^{\mathrm{rem}} &\geq k,\\
T_t^{\mathrm{rem}} &\geq k,
\end{aligned}
\label{eq:repair_action_mask}
\end{equation}
where $d_t$ is the estimated DoF deficit,
$B_t^{\mathrm{rem}}$ is the remaining repair
budget, and $T_t^{\mathrm{rem}}$ is the number
of remaining transmission opportunities.

These conditions ensure that repair is considered
only when a decoding deficit exists and sufficient
resources are available to transmit the selected
repair burst. Similarly, the source action is
available only when unsent source packets and
transmission opportunities remain.

The action mask restricts the scheduler to the
feasible action set $\mathcal{A}_t$. As a result,
the repair budget and deadline constraints are
enforced directly, allowing the learned policy
to focus on determining when repair transmission
is more beneficial than continuing source delivery.

\subsubsection{Reward Design}

Our objective is to maximize the KV utility recovered
before the decode-start deadline. However, block-level
utility is obtained only upon complete recovery,
resulting in sparse rewards. To provide intermediate
feedback, we introduce an auxiliary potential
\begin{equation}
\begin{aligned}
F_t
={}&
\lambda_{\mathrm{p}}
\frac{k_t^{\mathrm{dec}}}{N}
-
\lambda_{\mathrm{d}}
\frac{d_t}{\max\{1,d_t+B_t^{\mathrm{rem}}\}}
\\
&-
\lambda_{\mathrm{w}}
\frac{\log(1+W_t)}{\log(1+N)},
\end{aligned}
\label{eq:reward_potential}
\end{equation}
where $k_t^{\mathrm{dec}}$ denotes cumulative
decoding progress, $N$ is the total number of
source packets, and $W_t$ is the unacknowledged
coding-window size. The three terms encourage
decoding progress, penalize the DoF deficit relative
to the remaining repair budget, and discourage
excessive window growth, respectively.
We set $\lambda_{\mathrm{p}}=0.20$,
$\lambda_{\mathrm{d}}=0.15$, and
$\lambda_{\mathrm{w}}=0.05$.

The immediate reward is defined as
\begin{equation}
r_t =
J_{t+1}^{\mathrm{KV}}-J_t^{\mathrm{KV}}
+\widetilde{F}_{t+1}-\widetilde{F}_t,
\label{eq:reward_function}
\end{equation}
where $\widetilde{F}_t=F_t$ for nonterminal
states and is set to zero at termination.
With $\gamma=1$, the cumulative reward satisfies
\begin{equation}
\sum_{t=0}^{t_{\mathrm{end}}-1}r_t
=
J_{t_{\mathrm{end}}}^{\mathrm{KV}}
-J_0^{\mathrm{KV}}-F_0.
\label{eq:reward_telescope}
\end{equation}
Since the initial terms are independent of the
scheduling policy, maximizing the expected
cumulative reward is equivalent to maximizing
the expected KV utility at the deadline.
Thus, the auxiliary potential provides denser
training feedback without altering the original
optimization objective.


\subsection{Double-DQN-Based Repair Scheduling}
\label{subsec:ddqn_solution}

To solve the proposed MDP with continuous state
features and discrete actions, we employ Double
Deep Q-Network (Double DQN) to learn the
source--repair scheduling policy. The Q-network
takes the 15-dimensional state as input and
consists of two fully connected hidden layers
with 64 ReLU units each, followed by four
outputs corresponding to the scheduling actions.

During offline training, an $\epsilon$-greedy
policy selects actions from the feasible action
set $\mathcal{A}_t$. The resulting transitions,
including the next-state action mask and terminal
indicator $z_t$, are stored in a replay buffer
for mini-batch updates. Double DQN uses the
online network $Q_\theta$ to select the next
feasible action:
\begin{equation}
u_{t+1}^{*}
=
\arg\max_{u\in\mathcal{A}_{t+1}}
Q_\theta(\mathbf{x}_{t+1},u),
\label{eq:double_dqn_selection}
\end{equation}
while the target network $Q_{\bar{\theta}}$
evaluates this action to construct the target
\begin{equation}
y_t
=
r_t+(1-z_t)\gamma
Q_{\bar{\theta}}(\mathbf{x}_{t+1},u_{t+1}^{*}),
\label{eq:double_dqn_target}
\end{equation}
where $\gamma=1$ and $z_t=1$ indicates a
terminal transition. For terminal transitions,
the target reduces to $y_t=r_t$.

The online network is trained by minimizing
the Huber loss
\begin{equation}
\mathcal{L}(\theta)
=
\frac{1}{|\mathcal{M}|}
\sum_{i\in\mathcal{M}}
\ell_{\mathrm{Huber}}
\left(Q_\theta(\mathbf{x}_i,u_i)-y_i\right),
\label{eq:ddqn_loss}
\end{equation}
where $\mathcal{M}$ denotes a sampled mini-batch.
We use Adam for parameter updates and periodically
synchronize the target network with the online
network. The replay buffer is initially populated
with trajectories from a deficit-based heuristic
to facilitate early exploration, without imposing
any imitation-learning objective.

After training, only the online Q-network is
deployed at the sender. At each decision epoch,
the scheduler selects the highest-valued feasible
action:
\begin{equation}
u_t^{*}
=
\arg\max_{u\in\mathcal{A}_t}
Q_\theta(\mathbf{x}_t,u).
\label{eq:online_action_selection}
\end{equation}
The selected action determines whether to continue
utility-ordered source transmission or generate
a repair burst using ACK-sliding Streaming FEC.
Thus, online scheduling requires only a single
lightweight network inference per decision epoch.

\section{Simulation and Evaluation}
\label{sec:evaluation}

We evaluate PriorityFEC to examine two aspects:
(i) whether utility-aware source ordering improves
the recovery of important KV information before
the decoding deadline, and
(ii) whether the recovered KV cache improves downstream
LLM task performance.
We first introduce the experimental setup and
implementation details, followed by evaluations of
KV recovery, task quality, component contributions,
and computational overhead.

\subsection{Experimental Setup}
\label{subsec:eval_setup}

\textbf{Models and workloads.}
We evaluate PriorityFEC using two open-source LLMs,
Qwen2.5-7B-Instruct and Llama-3.1-8B-Instruct.
Three question-answering datasets from LongBench-E~\cite{bai2024longbench}
are considered, including \textsc{2WikiMultiHopQA},
\textsc{HotpotQA}, and \textsc{TriviaQA}.
These tasks require the models to retrieve and reason
over information from long input contexts.
We limit the input context to 1,536 tokens and
use the same preprocessing settings for all
compared schemes.
Inference quality is evaluated using token-level F1.

For each model--dataset pair, we select 48 examples
and evaluate each example under five paired network
realizations, resulting in 240 evaluations per
operating point.
All compared schemes use the same input examples
and packet-loss realizations to ensure fair comparisons.

\textbf{Network and deadline configuration.}
Unless otherwise specified, we consider a
100-Mbps bottleneck link with a 20-ms round-trip
time (RTT).
Packet losses are modeled using a two-state
Gilbert--Elliott (GE) channel~\cite{gilbert1960burst,elliott1963errors} with parameters
$(\epsilon_G,\epsilon_B,p_{GB},p_{BG})
=(0.001,0.50,0.02,0.10)$,
corresponding to an average packet-loss rate
of approximately $8.42\%$.

To characterize the available transmission time,
we define the normalized deadline ratio as
\begin{equation}
\rho
\triangleq
\frac{D-T_{\mathrm{ow}}}{T_{\mathrm{ser}}},
\qquad
T_{\mathrm{ser}}=\frac{S_{\mathrm{KV}}}{C},
\label{eq:deadline_ratio_eval}
\end{equation}
where $D$ is the decoding deadline,
$T_{\mathrm{ow}}$ is the one-way propagation delay,
$S_{\mathrm{KV}}$ is the complete KV-cache size,
and $C$ is the bottleneck link rate.
$T_{\mathrm{ser}}$ denotes the time required to
transmit the complete KV cache over a loss-free link.

\textbf{Compared schemes.}
We compare PriorityFEC against the following
transport schemes:
\begin{itemize}
    \item \textbf{TCP and QUIC:}
    Reliable transport protocols that recover
    lost packets through retransmissions.

    \item \textbf{UDP:}
    Transmits KV packets in their original
    FIFO order without retransmission or FEC.

\item \textbf{Uniform FEC:} This baseline transmits KV packets in FIFO order and applies dense GF(256) RLNC to independent blocks of 32 source packets. Repair packets are sent after each source block with a fixed repair-to-source ratio of 10\%, without ACK-driven adaptation or utility-aware scheduling.

    \item \textbf{FlEC-RT:}
    A real-time variant of FlEC~\cite{michel2023flec}
    that uses GF$(2^8)$ random linear coding
    over a moving coding window.
    It adaptively schedules repair packets based on
    receiver feedback, estimated channel conditions,
    and the remaining deadline, without considering
    model-derived KV utility.

    \item \textbf{Priority+Fixed StreamingFEC}
    The proposed scheme that combines utility-aware
    source ordering with a fixed 10\% repair schedule
    Streaming FEC.    
    \item \textbf{PriorityFEC:}
    The proposed scheme that combines utility-aware
    source ordering with RL-based adaptive
    Streaming FEC.
\end{itemize}

\textbf{Evaluation metrics.}
We consider two performance metrics.
The first is \emph{task score}, measured by the
token-level F1 obtained from the reconstructed
KV cache. The second is \emph{recovered KV utility}, defined
as the normalized utility of KV blocks completely
recovered before the decoding deadline.

These metrics respectively evaluate the effectiveness
of KV transmission and its impact on LLM inference.

\subsection{Implementation and Training Details}
\label{subsec:implementation_training}

\textbf{Prototype implementation.}
We implement the LLM inference pipeline using
PyTorch~\cite{paszke2019pytorch} and Hugging Face
Transformers~\cite{wolf2020transformers}, with a
native C implementation of Streaming FEC over
GF$(2^8)$. The prototype employs two NVIDIA
Jetson AGX Orin 64GB edge platforms.
For a batch size of one, the BF16 key and value
tensors have shape $[1,H_{\mathrm{KV}},T,d_h]$.
Each transport block contains the KV packets of
16 consecutive tokens from one layer.
We use 1,400-B application frames, of which
1,368 B carry KV data and the remainder
contains coding and fragmentation metadata.

\textbf{Partial-KV reconstruction.}
Recovered fragments are restored to their
original tensor positions. In the default
zero-fill scheme, a BF16 element is retained
only if both bytes are recovered; otherwise,
it is set to zero. The reconstructed KV cache
preserves the original tensor dimensions and
token positions, enabling decode-only inference
without repeating prefill.
We also implement a mask-based alternative
that excludes incomplete KV head--token rows
from attention, while preserving the original
RoPE positions and newly generated KV information.
Unless otherwise specified, all results use
zero-fill reconstruction.

\textbf{Offline scheduler training.}
The scheduler employs a lightweight Double DQN
with a $15$--$64$--$64$--$4$ architecture and
5,444 trainable parameters. It is trained
offline using synthetic utility-ranked KV blocks
and randomized Gilbert--Elliott (GE) channels,
without accessing QA samples or task-level
evaluation scores. Training uses an approximate
sparse-coding model, whereas evaluation employs
the actual GF$(2^8)$ codec.

Following initialization with randomized
transport trajectories, we conduct 120 additional
training episodes using Adam~\cite{kingma2015adam}
with a learning rate of $10^{-4}$ and
$\gamma=1$. The exploration rate decreases
from 0.25 to 0.05, with gradient clipping
at a maximum norm of five. Network updates
occur every 16 transitions, and the target
network is synchronized every 500 gradient
steps. Training comprises 321,032 transitions
and 16,550 gradient updates. The resulting
policy remains fixed throughout evaluation.

\textbf{Evaluation scope.}
Training covers link rates of 50 and 100 Mbps
and RTTs of 14--26 ms. We additionally evaluate
the fixed policy using a recorded 5G loss trace
with an assumed service rate of 20.48 Mbps,
without retraining. This setting examines
transfer to a measured loss pattern outside
the training bandwidth range, rather than
establishing generalization across arbitrary
network conditions.

The evaluation also involves a change in
packetization scale between training and
deployment. Furthermore, the current attention
profiling implementation constructs full
attention tensors before extracting the last
64 query positions, limiting its scalability
to longer contexts. Consequently, scalability
to 4K/8K contexts and broader generalization
under unseen network conditions remain
outside the scope of this evaluation.

\subsection{End-to-End Task Quality}
\label{subsec:task_performance}

\begin{figure*}[t]
    \centering
    \subfloat[Qwen2.5-7B, 2WikiMultiHopQA.]{
        \includegraphics[width=0.32\textwidth]
        {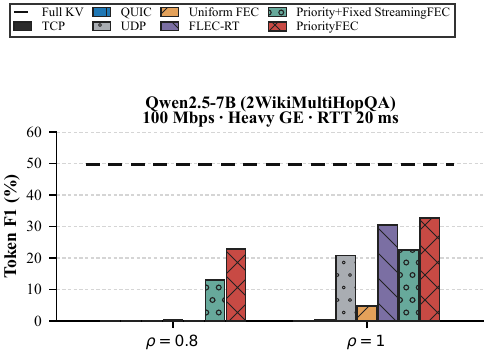}}
    \hfill
    \subfloat[Qwen2.5-7B, HotpotQA.]{
        \includegraphics[width=0.32\textwidth]
        {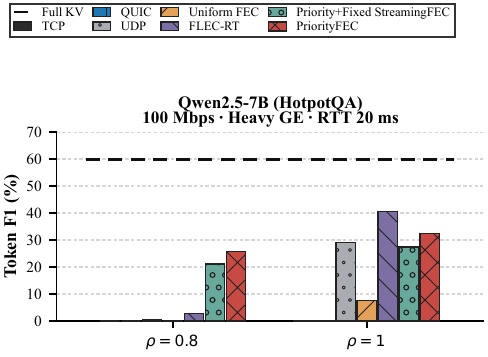}}
    \hfill
    \subfloat[Qwen2.5-7B, TriviaQA.]{
        \includegraphics[width=0.32\textwidth]
        {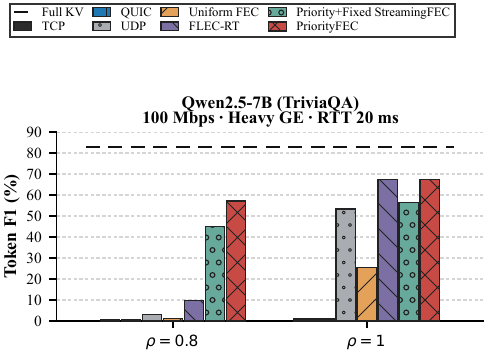}}
    \\
    \subfloat[Llama-3.1-8B, 2WikiMultiHopQA.]{
        \includegraphics[width=0.32\textwidth]
        {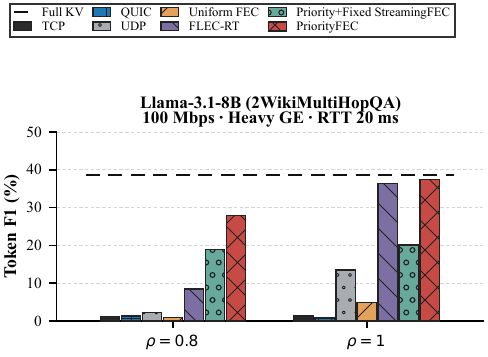}}
    \hfill
    \subfloat[Llama-3.1-8B, HotpotQA.]{
        \includegraphics[width=0.32\textwidth]
        {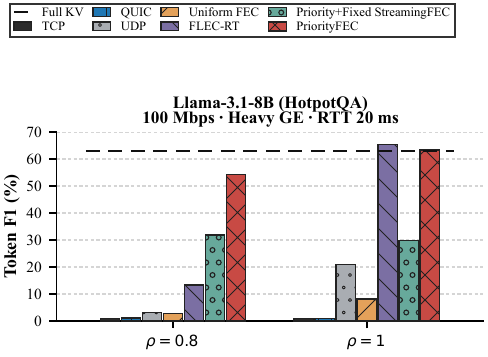}}
    \hfill
    \subfloat[Llama-3.1-8B, TriviaQA.]{
        \includegraphics[width=0.32\textwidth]
        {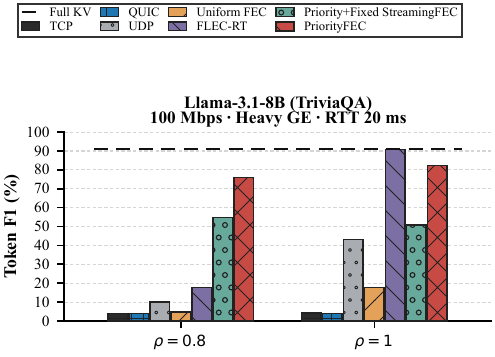}}

    \caption{End-to-end token-level F1 under different decode-start
    deadlines on the 100-Mbps burst-loss GE link.
    PriorityFEC provides substantial gains under $\rho=0.8$, where
    complete loss-free KV serialization is infeasible.
    As the deadline is relaxed to $\rho=1.0$, the performance gap
    decreases because more KV cache can be recovered regardless of
    source position.}
    \label{fig:task_score}
\end{figure*}

Fig.~\ref{fig:task_score} evaluates the downstream token-level F1 obtained
when decoding starts from the KV cache available at the deadline.
The results demonstrate that our proposed scheme achieves substantial advantages under tight deadline budgets, effectively improving task accuracy.

At $\rho=0.8$, PriorityFEC
consistently outperforms all transport baselines
across the six model--workload combinations.
For Qwen2.5-7B, PriorityFEC achieves token F1
scores of approximately $22.3\%$, $25.1\%$,
and $56.6\%$ on 2WikiMultiHopQA, HotpotQA,
and TriviaQA, respectively, substantially
outperforming the competing schemes.
The advantage is particularly pronounced on
TriviaQA, where PriorityFEC preserves more than
half of the Full-KV task score, despite a deadline
budget corresponding to only $80\%$ of the
loss-free KV serialization time.

A similar but more pronounced trend is observed
for Llama-3.1-8B. PriorityFEC achieves token F1
scores of approximately $27.5\%$, $53.7\%$,
and $75.2\%$ on the three workloads, respectively,
compared with only $8.1\%$, $12.7\%$, and
$16.8\%$ achieved by FLEC-RT.
These substantial differences indicate that
adaptive loss recovery alone is insufficient
to maintain task quality when source packets
are transmitted without considering their
KV utility. Under tight deadlines, prioritizing
the delivery of high-utility KV blocks is
therefore critical to preserving inference quality.


Across all six model--workload combinations at $\rho=0.8$,
PriorityFEC achieves an average F1 score of approximately
43.4\%, compared with 8.1\% for FlEC-RT, corresponding
to an absolute improvement of 35.3 percentage points.
The Priority+Fixed StreamingFEC baseline also achieves
substantial improvements over utility-unaware transport
schemes, highlighting the importance of source ordering
under stringent deadlines. Nevertheless, PriorityFEC
consistently achieves the highest F1 scores across the
six evaluated cases, indicating that adaptive repair
scheduling can further improve downstream task quality
when transmission opportunities are limited.

As the deadline budget increases to $\rho=1.0$,
the performance gap between PriorityFEC and
the baselines becomes less pronounced. With
additional transmission opportunities, all
lossy-transport schemes can deliver and recover
more KV information before decoding.
For Qwen2.5-7B, FLEC-RT becomes more competitive
and outperforms PriorityFEC on HotpotQA.
For Llama-3.1-8B, PriorityFEC achieves F1
scores of approximately $37.1\%$ and $62.8\%$
on 2WikiMultiHopQA and HotpotQA, respectively,
approaching their corresponding Full-KV
references. Nevertheless, FLEC-RT also
surpasses PriorityFEC on Llama-3.1-8B TriviaQA,
indicating that the relative advantages of
utility-aware ordering diminish as the
deadline becomes less restrictive.

These results highlight the role of the deadline
constraint in determining the benefits of
utility-aware transmission. At $\rho=0.8$,
the available transmission time is insufficient
to deliver the complete KV cache even without
packet losses. Source-packet ordering therefore
strongly influences which KV blocks become
available before decoding begins. As $\rho$
increases, a larger portion of the KV cache
can be transmitted and recovered, reducing
the marginal benefit of prioritizing
high-utility blocks. In this regime, the
effectiveness of loss recovery becomes
increasingly important. Overall, PriorityFEC
provides its largest gains under stringent
deadline constraints, where preserving the
most valuable KV information is essential
to maintaining downstream inference quality.

\subsection{Deadline-Available KV Utility}
\label{subsec:recovered_utility}

In this part, we use recovered KV utility as the evaluation metric to assess whether the proposed scheduling scheme can achieve higher KV utility than alternative heuristic scheduling strategies.

\begin{figure*}[t]
    \centering
    \subfloat[Qwen2.5-7B, 2WikiMultiHopQA.]{
        \includegraphics[width=0.32\textwidth]
        {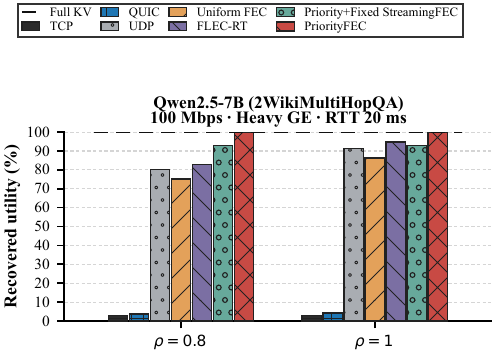}}
    \hfill
    \subfloat[Qwen2.5-7B, HotpotQA.]{
        \includegraphics[width=0.32\textwidth]
        {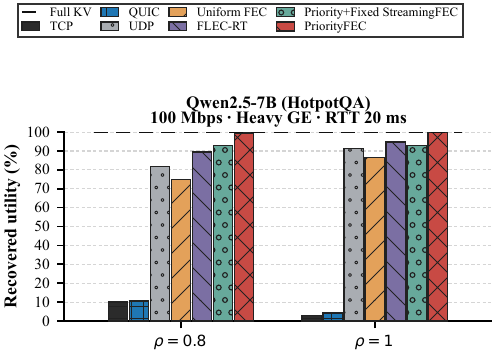}}
    \hfill
    \subfloat[Qwen2.5-7B, TriviaQA.]{
        \includegraphics[width=0.32\textwidth]
        {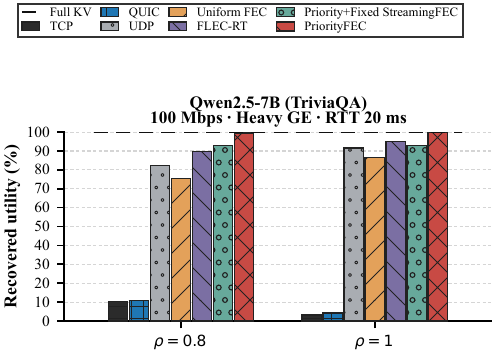}}
    \\
    \subfloat[Llama-3.1-8B, 2WikiMultiHopQA.]{
        \includegraphics[width=0.32\textwidth]
        {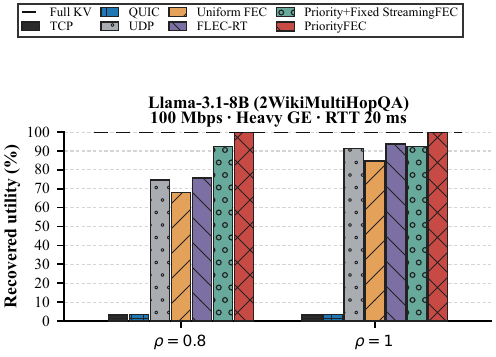}}
    \hfill
    \subfloat[Llama-3.1-8B, HotpotQA.]{
        \includegraphics[width=0.32\textwidth]
        {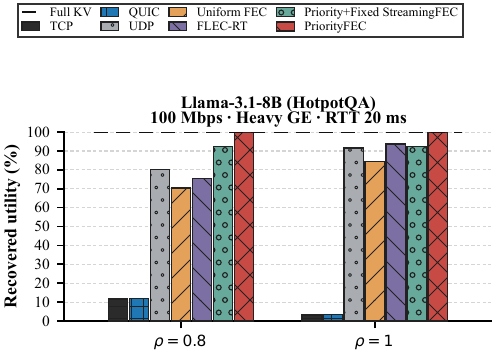}}
    \hfill
    \subfloat[Llama-3.1-8B, TriviaQA.]{
        \includegraphics[width=0.32\textwidth]
        {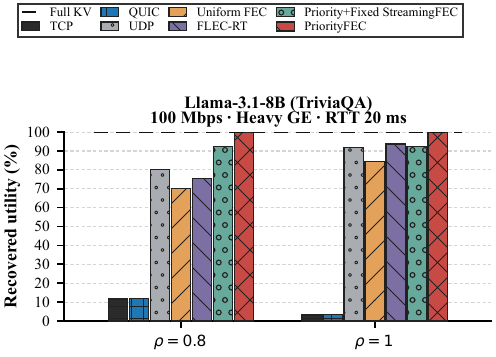}}
    \caption{Normalized KV utility recovered before the decode-start
    deadline on the 100-Mbps burst-loss GE link.
    PriorityFEC consistently preserves the highest-utility KV regions,
    with its advantage being most pronounced under the tight
    $\rho=0.8$ deadline.}
    \label{fig:recovered_utility}
\end{figure*}

As shown in Fig.~\ref{fig:recovered_utility},
PriorityFEC consistently achieves higher recovered
KV utility than the transport baselines, with the
largest advantages observed under the deadline
$\rho=0.8$ and $\rho=1.0$. For Qwen2.5-7B, shown in the upper
panels, PriorityFEC preserves approximately $99\%$
of the normalized KV utility across the three
workloads, substantially outperforming the competing
schemes under the deadline $\rho=0.8$. A similar trend is observed for
Llama-3.1-8B in the lower panels, demonstrating
the effectiveness of utility-aware transmission
across different LLM architectures and workloads.
Across all six model--workload combinations,
FlEC-RT retains
approximately $75$--$89\%$ of the KV utility.
On average, PriorityFEC preserves $99.1\%$,
compared with $82.3\%$ for FLEC-RT baseline, yielding an absolute
improvement of $16.8$ percentage points.

These results highlight the distinction between
\emph{byte delivery} and \emph{useful-state delivery}.
At $\rho=0.8$, the deadline is insufficient to
transmit the complete KV cache even without packet
losses. Under this constraint, utility-aware
ordering prioritizes the transmission of the most
valuable KV blocks, while adaptive repair
scheduling helps recover lost information.
By jointly considering source delivery and loss
recovery, PriorityFEC preserves a larger fraction
of the model-relevant KV information within the
available transmission budget.

As the deadline increases to $\rho=1.0$,
the performance gap becomes smaller for both
models. PriorityFEC maintains approximately
$99.2\%$ of the normalized KV utility, while
FLEC-RT achieves roughly $93$--$94\%$.
The average advantage over FLEC-RT
decreases to approximately $5.6$ percentage points.
This trend suggests that utility-aware ordering
is particularly beneficial when the transmission
budget is insufficient to deliver the entire KV
cache. With additional transmission opportunities,
more lower-priority KV blocks can be recovered,
reducing the relative importance of source ordering.

Overall, these results demonstrate that PriorityFEC
improves not merely the amount of recovered KV packets,
but more importantly, the utility of the KV information
available at the decode-start deadline. We next
investigate whether this transport-level advantage
translates into improved downstream task performance.

\subsection{Component Ablation}
\label{subsec:ablation}

To evaluate the contributions of FEC protection and
adaptive repair scheduling, we conduct an ablation
study using Qwen2.5-7B-Instruct on the three QA
workloads. All experiments use the same
100-Mbps Gilbert--Elliott channel, paired loss
realizations, and maximum repair budget as the
main evaluation. We consider three deadline ratios,
$\rho\in\{0.6,0.8,1.0\}$, and compare the following
variants:
\begin{itemize}
    \item \textbf{Priority-only:}
    Transmits KV blocks in utility-aware order
    without FEC protection.

    \item \textbf{Priority+FlEC scheduler:}
    Combines utility-aware source ordering with
    a feedback-driven heuristic repair scheduler deployed in streaming FEC.

    \item \textbf{PriorityFEC:}
    The complete scheme that combines
    utility-aware source ordering with the
    proposed Double-DQN-based repair scheduler.
\end{itemize}

\begin{figure*}[t]
    \centering
    \subfloat[HotpotQA.]{
        \includegraphics[width=0.32\textwidth]
        {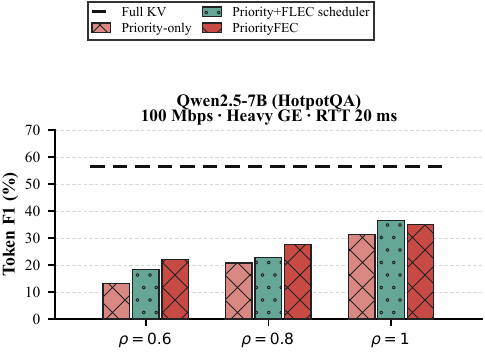}}
    \hfill
    \subfloat[TriviaQA.]{
        \includegraphics[width=0.32\textwidth]
        {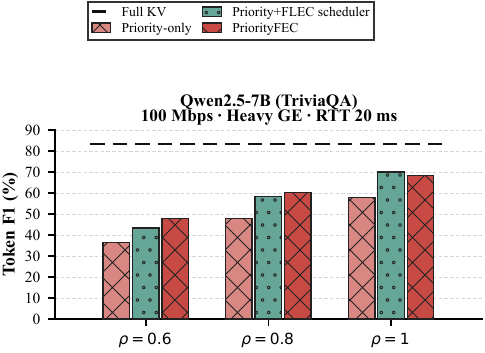}}
    \hfill
    \subfloat[2WikiMultiHopQA.]{
        \includegraphics[width=0.32\textwidth]
        {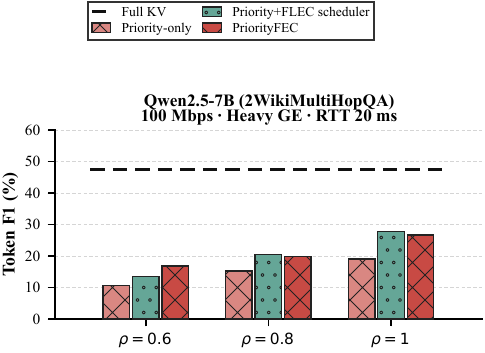}}
    \caption{Ablation results of PriorityFEC on
    Qwen2.5-7B-Instruct under the 100-Mbps
    heavy-GE channel.}
    \label{fig:qwen_ablation}
\end{figure*}

Fig.~\ref{fig:qwen_ablation} compares the
token-level F1 achieved by the three variants.
Across all workloads and deadline settings,
both FEC-based schemes outperform Priority-only,
demonstrating the benefit of packet-loss recovery
even when important KV blocks are transmitted first.
Without FEC protection, losses may prevent
high-priority KV blocks from being fully recovered
before decoding begins. Introducing repair
transmissions improves the availability of these
blocks and consequently benefits downstream
task performance.

The impact of learned repair scheduling is most
evident under stringent deadline constraints.
At $\rho=0.6$, PriorityFEC achieves the highest
F1 score across all three workloads, outperforming
both Priority-only and Priority+FlEC scheduler.
In this regime, source and repair transmissions
compete for a severely limited number of
transmission opportunities. By incorporating
the remaining deadline, decoding progress, and
KV utility into its scheduling decisions, the
learned policy provides a more effective
source--repair tradeoff than the heuristic
scheduler in the evaluated scenarios.

As the deadline becomes less restrictive,
the performance difference between the two
repair schedulers decreases. At $\rho=0.8$,
PriorityFEC maintains an advantage on HotpotQA
and TriviaQA while achieving comparable
performance on 2WikiMultiHopQA. At $\rho=1.0$,
Priority+FlEC scheduler slightly outperforms
PriorityFEC across all three workloads.
These results indicate that the benefit of
learned repair scheduling is operating-regime
dependent rather than universal.

Overall, the ablation results establish two
complementary observations. First, FEC-based
loss recovery substantially improves the
effectiveness of utility-aware KV transmission.
Second, learned source--repair scheduling provides
additional task-quality gains when transmission
opportunities are severely constrained, whereas
its advantage over heuristic scheduling diminishes
under more relaxed deadlines.

\subsection{Runtime Overhead and Computational Feasibility}
\label{subsec:runtime_overhead}

PriorityFEC requires online coding and scheduling
decisions to keep pace with packet transmission.
To evaluate its computational feasibility, we
measure four components: state construction and
feedback processing, Double-DQN inference,
GF$(2^8)$ encoding, and GF$(2^8)$ decoding.
The measurements cover all three QA workloads
and their paired network realizations.
Table~\ref{tab:runtime_overhead} reports the
processing overhead per 1,000 transmitted packets,
together with the latency of each online
scheduling decision.

\begin{table*}[t]
\centering
\caption{Measured runtime overhead of PriorityFEC.
Component overhead is normalized per 1,000
transmitted packets. Control latency is measured
per scheduling decision.}
\label{tab:runtime_overhead}
\scriptsize
\setlength{\tabcolsep}{4.5pt}
\renewcommand{\arraystretch}{1.08}
\begin{tabular}{lcccccccc}
\toprule
Model & $\rho$
& State/FB
& DDQN
& Enc.
& Dec.
& Total
& Control Mean/P99
& Overrun \\
& & \multicolumn{5}{c}{(ms per 1k packets)}
& ($\mu$s)
& (\%) \\
\midrule
Qwen2.5-7B
& 0.8
& 9.42 & 3.41 & 10.31 & 7.19
& 30.34
& 75.98 / 94.35
& 0.009 \\

Qwen2.5-7B
& 1.0
& 9.28 & 3.36 & 11.14 & 8.76
& 32.53
& 76.99 / 95.67
& 0.011 \\

Llama-3.1-8B
& 0.8
& 9.38 & 3.43 & 10.32 & 7.29
& 30.41
& 76.53 / 95.18
& 0.019 \\

Llama-3.1-8B
& 1.0
& 9.28 & 3.33 & 11.34 & 9.46
& 33.41
& 78.00 / 97.58
& 0.016 \\
\bottomrule
\end{tabular}
\end{table*}

As shown in Table~\ref{tab:runtime_overhead},
the total processing overhead ranges from
30.34 to 33.41 ms per 1,000 packets, with
similar results across the two LLMs.
At the evaluated link rate of 100 Mbps,
transmitting a 1,400-B packet requires
112 $\mu$s, or 112 ms per 1,000 packets.
The measured processing overhead is therefore
equivalent to approximately 27--30\% of the
packet serialization time, indicating that
the implementation has sufficient average
computational capacity for the evaluated
transmission rate.

The online scheduler also introduces limited
decision latency. Its mean execution time
ranges from 75.98 to 78.00 $\mu$s, while
the P99 latency remains below 98 $\mu$s
across all settings.
Both are lower than the 112-$\mu$s packet
serialization interval.
Moreover, only 0.009--0.019\% of scheduling
decisions exceed this interval.
These results indicate that the adaptive
scheduler can generally make source--repair
decisions within the available packet
transmission time.

The component breakdown further shows that
Double-DQN inference requires only
3.33--3.43 ms per 1,000 packets, accounting
for a small fraction of the total overhead.
Most computation is spent on packet encoding,
decoding, and feedback processing rather than
neural network inference.
The similar overhead observed for the two
LLMs also suggests that the transport-side
processing cost is primarily determined by
packet and coding operations under the
evaluated configurations.
Overall, these measurements demonstrate
the computational feasibility of PriorityFEC
at the tested link rate, although
end-to-end timing performance in a fully
distributed deployment requires further
validation.

\subsection{Measured 5G Trace Replay}
\label{subsec:5g_trace_replay}

To further evaluate PriorityFEC under realistic
packet-loss conditions, we conduct trace-driven
experiments using the public 5G Campus Networks
dataset~\cite{Rischke2021campus}.
We select a 300-s standalone (SA) uplink UDP-Ping
trace with an average packet-loss rate of 11.65\%
and set the replay transmission rate to
20.48 Mbps, matching the recorded offered rate.
Packet losses are reproduced according to their
observed timing in the trace.
The first 30 s are used for channel-history
calibration, and evaluation windows are selected
from the remaining segment.
The forward and reverse propagation delays are
fixed at 9.36 and 13.20 ms, respectively,
corresponding to an RTT of approximately 22.6 ms.
The feedback channel is assumed reliable,
and packet-level delay variations are not replayed.
Since the original trace uses 256-B packets,
we apply its time-dependent loss pattern
to the 1,400-B KV packets.

We retain the two LLMs, three QA workloads,
and deadline ratios $\rho\in\{0.8,1.0\}$.
For each model--workload pair, we evaluate
eight examples under two paired replay windows,
resulting in 16 evaluations per operating point.
All compared schemes use the same inputs,
loss traces, and transmission constraints
within each paired experiment.

\begin{figure*}[t]
    \centering
    \subfloat[Qwen2.5-7B, 2WikiMultiHopQA.]{
        \includegraphics[width=0.32\textwidth]
        {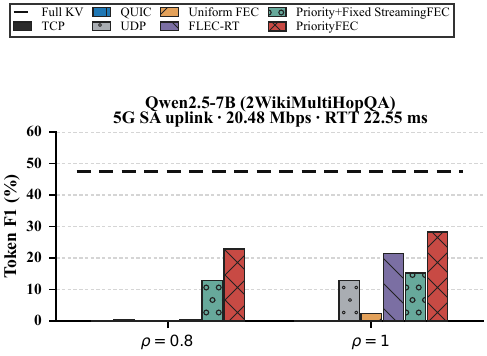}}
    \hfill
    \subfloat[Qwen2.5-7B, HotpotQA.]{
        \includegraphics[width=0.32\textwidth]
        {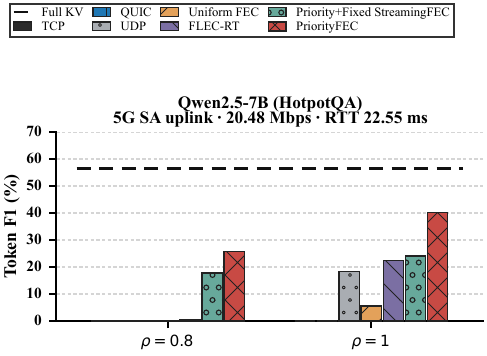}}
    \hfill
    \subfloat[Qwen2.5-7B, TriviaQA.]{
        \includegraphics[width=0.32\textwidth]
        {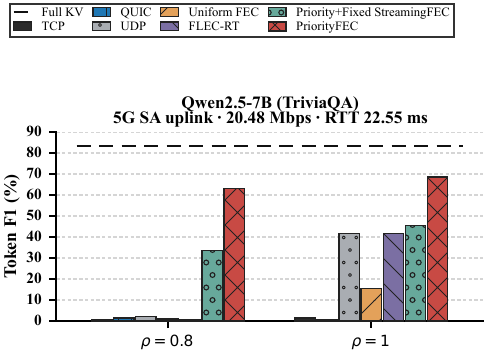}}
    \\
    \subfloat[Llama-3.1-8B, 2WikiMultiHopQA.]{
        \includegraphics[width=0.32\textwidth]
        {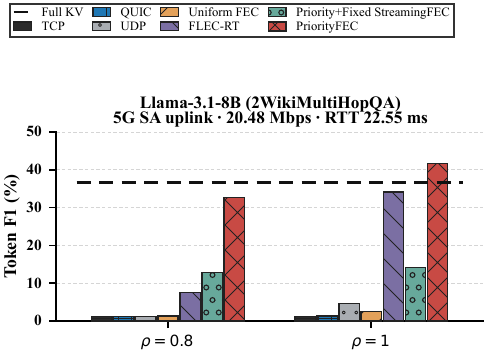}}
    \hfill
    \subfloat[Llama-3.1-8B, HotpotQA.]{
        \includegraphics[width=0.32\textwidth]
        {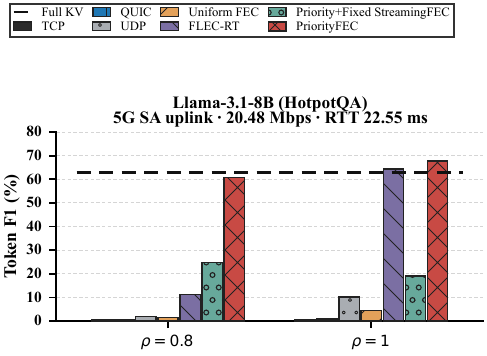}}
    \hfill
    \subfloat[Llama-3.1-8B, TriviaQA.]{
        \includegraphics[width=0.32\textwidth]
        {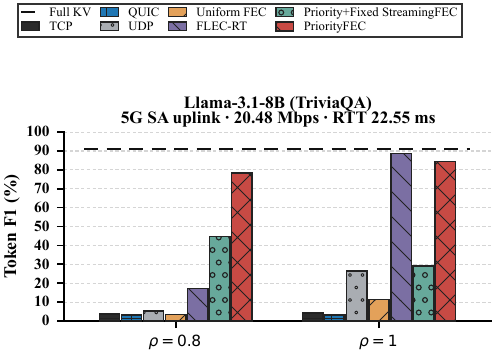}}
    \caption{Token-level F1 under measured 5G SA
    uplink loss replay. Results are averaged over
    eight examples and two paired replay windows
    per example.}
    \label{fig:5g_task_score}
\end{figure*}

Fig.~\ref{fig:5g_task_score} compares the
token-level F1 achieved by different transport
schemes.
Under the stringent deadline of $\rho=0.8$,
PriorityFEC consistently outperforms the five
main transport baselines across all six
model--workload combinations.
For Qwen2.5-7B, PriorityFEC achieves F1 scores
of 36.6\%, 28.2\%, and 75.8\% on
2WikiMultiHopQA, HotpotQA, and TriviaQA,
respectively, compared with 0.0\%, 2.1\%,
and 1.8\% for FlEC-RT.
Similar improvements are observed for
Llama-3.1-8B, where PriorityFEC achieves
40.0\%, 85.0\%, and 95.8\%, compared with
5.3\%, 18.7\%, and 20.1\% for FlEC-RT.
These results indicate that utility-aware
transmission and adaptive repair scheduling
can help maintain LLM task performance
under the measured burst-loss conditions.

When the deadline is relaxed to $\rho=1.0$,
PriorityFEC remains competitive, although its
advantage is not consistent across all workloads.
For example, FlEC-RT achieves an F1 score of
81.0\% on Llama-3.1-8B HotpotQA, slightly
exceeding the 79.7\% obtained by PriorityFEC.
We also observe that the F1 score of PriorityFEC
on Llama-3.1-8B 2WikiMultiHopQA decreases from
40.0\% to 25.7\% as $\rho$ increases from
0.8 to 1.0. This observation indicates that
improvements in transmission conditions do
not necessarily translate into monotonic gains
in downstream task performance, since the
generated responses also depend on the specific
KV cache reconstructed at decode startup.

Overall, the trace-driven results show that
PriorityFEC can improve downstream LLM task
performance under the selected 5G loss pattern,
particularly when the transmission deadline
is stringent.
However, these findings are based on one
measured trace and a limited number of QA
examples. Further evaluations across diverse
5G conditions and larger workloads are needed
to establish broader performance generality.

It is worth noting that the scheduler directly
optimizes deadline-available KV utility rather than
downstream task F1.
While recovered KV utility provides a transport-level
measure of useful KV information availability, the resulting
task quality also depends on interactions among
recovered KV regions during model inference.
Consequently, an improvement in recovered utility
does not necessarily translate into a proportional
or monotonic improvement in F1.

\section{Conclusion and Limitation}
\label{sec:conclusion}

In this paper, we have proposed PriorityFEC, a
utility-aware Streaming FEC scheme for deadline-constrained
KV-cache handoff in collaborative edge LLM systems.
PriorityFEC aims to maximize the utility of KV information
available to the receiving agent before the decoding
deadline, rather than requiring complete KV-cache delivery.
To achieve this objective, we first develop a utility-aware
source ordering strategy that prioritizes important KV
blocks for transmission. We then design an adaptive
Streaming FEC scheme with a lightweight Double-DQN
scheduler to dynamically allocate repair transmissions
based on network feedback and KV utility.
Experiments using Qwen2.5-7B-Instruct and
Llama-3.1-8B-Instruct demonstrate that PriorityFEC
improves recovered KV utility and downstream task
quality, particularly under stringent deadlines and
bursty packet losses. Ablation experiments further demonstrate that
the learned repair scheduler improves downstream
task quality over heuristic scheduling under
stringent transmission deadlines, while its
advantage diminishes as more transmission
opportunities become available.
These results demonstrate the effectiveness of
utility-aware transport in supporting timely KV-cache
handoff among collaborative edge agents.

Although this work incorporates KV-cache utility into
transmission scheduling, it does not jointly optimize
KV-cache compression, eviction, and network transport.
In future work, we will explore the co-design of these
mechanisms to further reduce transmission overhead
while preserving task-relevant KV information.
Such integration may enable more efficient and
adaptive KV-cache transport protocols, facilitating
low-latency context sharing and handoff in edge
multi-agent systems.

\bibliographystyle{IEEEtran}
\bibliography{reference}
\vfill
\end{document}